# Two-photon vibrational excitation of air by long-wave infrared laser pulses

J.P. Palastro, J. Peñano, L.A. Johnson, and B. Hafizi
*Naval Research Laboratory, Washington DC 20375-5346, USA*

J.K. Wahlstrand, and H.M. Milchberg
*Institute for Research in Electronics and Applied Physics*
*University of Maryland, College Park MD 20740, USA*

**Abstract**

Ultrashort long-wave infrared (LWIR) laser pulses can resonantly excite vibrations in $N_2$ and $O_2$ through a two-photon transition. The absorptive, vibrational component of the ultrafast optical nonlinearity grows in time, starting smaller than, but quickly surpassing, the electronic, rotational, and vibrational refractive components. The growth of the vibrational component results in a novel mechanism of $3^{rd}$ harmonic generation, providing an additional two-photon excitation channel, fundamental + $3^{rd}$ harmonic. The original and emergent two-photon excitations drive the resonance exactly out of phase, causing spatial decay of the absorptive, vibrational nonlinearity. This nearly eliminates two-photon vibrational absorption. Here we present simulations and analytical calculations demonstrating how these processes modify the ultrafast optical nonlinearity in air. The results reveal nonlinear optical phenomena unique to the LWIR regime of ultrashort pulse atmospheric propagation.

# I. Introduction

Ultrashort laser pulses propagating through atmosphere drive an ultrafast dielectric response by nonlinearly polarizing the constituent atoms and molecules [1-3]. For the primary constituents, the diatomic molecules $N_2$ and $O_2$, the response consists of three motions: electronic, rotational, and vibrational. The time scale for each of these varies substantially, and, when compared with the pulse duration or frequency, determines their dynamic contribution to the dielectric response.

Bound electron dynamics occur on attosecond time scales, $\tau_e \sim \hbar U_I^{-1}$ where $U_I$ is the ionization potential. The electrons, as a result, respond nearly instantaneously to ultrashort (~fs to ps) laser pulses with near-ultraviolet or longer periods (> fs). In contrast, the rotational and vibrational dynamics involve motion of the nuclei, and occur on much longer time scales. The rotational motion has a characteristic time scale of picoseconds, $\tau_r \sim \hbar^{-1} I_M$ where $I_M$ is the moment of inertia [4-7]. Pulses with durations exceeding this adiabatically align the molecules and experience a near-instantaneous rotational response [4-7]. Much shorter pulses, on the other hand, impulsively align the molecules and experience a delayed rotational response [4-7]. A similar transition, from an instantaneous to delayed response, occurs during vibrational excitations but at much shorter pulse durations [8-12]. The vibrational motion occurs on the few femtosecond timescale, $\tau_v \sim \Omega_v^{-1}$ where $\Omega_v$ is the vibrational frequency, and thus, near-infrared pulses must be close to single-cycle for impulsive excitation [8-12].

While a non-adiabatic vibrational excitation can be achieved with few-cycle pulses, there is another option: using a longer wavelength pulse. Neither $N_2$ nor $O_2$ possess a permanent dipole moment; the laser pulse driven rotation and vibration is mediated, instead, by an induced dipole. At the quantum level, the induced dipole interaction corresponds to a two-photon process, by which the laser pulse excites coherence between two (selection rule allowed) rotational or vibrational states. For pulses with frequencies incommensurate with the states' transition frequency, the pulse duration determines the dynamics, as described above. However, when the frequency equals half the transition frequency, the pulse resonantly excites coherence between the states. Specifically, long wave infrared (LWIR) pulses with wavelengths near $\lambda = 4\pi c/\Omega_v$, 8.4 $\mu$m or 12.6 $\mu$m for $N_2$ or $O_2$ respectively [13], will resonantly excite vibrational coherence.

Even with the recent popularity of mid-infrared (MIR) ultrashort pulse propagation studies [14-18], the effect of this excitation on propagation, either on or off resonance, has yet to be examined.

Here we consider the ultrafast, two-photon vibrational excitation of $N_2$ and $O_2$, and examine its effect on LWIR pulse propagation. We find that the absorptive, vibrational component of the ultrafast optical nonlinearity grows in time, starting smaller than, but quickly surpassing, the electronic, rotational, and vibrational refractive components. The growth of the vibrational component results in a novel mechanism of accelerated 3$^{rd}$ harmonic generation. The resulting 3$^{rd}$ harmonic provides an additional resonant two-photon excitation channel, fundamental + 3$^{rd}$ harmonic. The original and emergent two-photon excitations drive the vibrational resonance exactly out of phase with each other, causing spatial decay of the absorptive vibrational nonlinearity. That is: co-propagation of the fundamental and the resonantly generated 3$^{rd}$ harmonic nearly eliminates two-photon vibrational absorption. Both simulations and analytical calculations are presented demonstrating how these processes modify the ultrafast optical nonlinearity in air.

## II. Ultrafast Optical Nonlinearity of Air

The nonlinear polarization density induced by laser pulses propagating through the atmosphere, $P = P_e + P_r + P_v$, includes contributions from the electronic, $P_e$, rotational, $P_r$, and vibrational, $P_v$, responses of (predominately) $N_2$ and $O_2$ molecules. Unless otherwise stated, a sum over the $N_2$ and $O_2$ polarizations, weighted by number density, is implied. Bound electrons respond near-instantaneously to LWIR pulses, such that the electronic polarization can be expressed as $P_e = \varepsilon_0 \chi_e E$ where

$$\chi_e = 4c\varepsilon_0 n_2 E^2 / 3 \quad (1)$$

is the electronic susceptibility, $E$ the transverse electric field of a linearly polarized laser pulse, and $n_2$ the second order nonlinear refractive index [19]. The values of $n_2$ and other parameters required for calculating the polarizations are summarized in Table 1.

The rotational and vibrational polarizations are derived using density matrix theory. Here we summarize the derivation and refer the reader to Refs. [5,12,21,22] for additional details. The molecules are modeled as spring-bound atom pairs with an anisotropic polarizability that varies with atomic separation. This allows a laser pulse to both align and stretch the molecules. The calculation involves expanding the density matrix in orders of the field-dependent potential responsible for the alignment and stretching. To lowest order in the expansion, alignment or stretching corresponds to excitation of coherence between two rotational or vibrational states respectively. At this order, no change in the state populations occurs.

In the absence of a field, the rovibrational energy is given by $U_{nj} = \hbar\Omega_v(n+\frac{1}{2}) + \frac{1}{2}\hbar^2 I_M^{-1} j(j+1)$, where $n$ and $j$ are the vibrational and total angular momentum quantum numbers respectively, $\Omega_v$ is the vibration frequency, and $I_M$ the moment of inertia. Assuming thermodynamic equilibrium, the zeroth order density matrix is then given by $\rho^0_{njm,jm} = \delta_{njm,njm} Z_p^{-1} D_j \exp[-U_{nj}/T]$, where $m$ is the angular momentum quantum number along the pulse polarization axis, $T$ is the temperature, $D_j$ a degeneracy factor associated with nuclear spin, and $Z_p = \sum_{nj}(2j+1)D_j \exp[-U_{nj}/T]$ is the partition function. A linear polarized laser pulse drives coherence between rotational or vibrational states separated by $\Delta j = \pm 2$ or $\Delta n = \pm 1$ respectively. The resulting rotational polarization can be expressed as $P_r = \varepsilon_0 \sum_j \chi_{r,j} E$, where

$$\left(\frac{\partial^2}{\partial t^2} + \omega_{j,j-2}^2\right)\chi_{r,j} = -\frac{32\pi^2 \varepsilon_0 \eta (\Delta\alpha)^2}{15 I_M} j(j-1) \left(\frac{\rho^0_{j,j}}{2j+1} - \frac{\rho^0_{j-2,j-2}}{2j-3}\right) E^2, \quad (2)$$

$\omega_{j,j-2} = \hbar I_M^{-1}(2j-1)$, $\eta$ is the number density, $\Delta\alpha$ the polarizability anisotropy, and $\rho^0_{j,j} = \sum_{nm} \rho^0_{njm,njm}$. Similarly, the vibrational polarization can be expressed as $P_v = \varepsilon_0 \chi_v E$, where

$$\left(\frac{\partial^2}{\partial t^2} + \Omega_v^2\right)\chi_v = \frac{4\pi^2 \varepsilon_0 \eta}{\mu}\left(\frac{\partial\alpha}{\partial Q}\right)^2 E^2, \quad (3)$$

$\mu$ is the reduced atomic mass, and $\partial\alpha/\partial Q$ the change in isotropic polarizability with atomic separation.

The full rovibrational response has been simplified in Eqs. (2) and (3) by applying the following observations. First, the molecules largely populate the ground vibrational state at atmospheric temperatures. Second, contributions proportional to $\partial\Delta\alpha/\partial Q$, including simultaneous rotational-vibrational excitations, contain factors making them an order of magnitude smaller than terms proportional to $\partial\alpha/\partial Q$. Finally, the time between geometric cross-section based collisions far exceeds the excitation times of interest.

## III. Two-Photon Vibrational Resonance

Equations (2) and (3) admit resonant solutions: When the period of $E^2$ is commensurate with the oscillator period, the molecular susceptibility undergoes temporal growth. The rotational resonances reside in the THz range, accessible by beating two laser frequencies together, or by appropriately delaying optical pulses [22]. Of interest here is the two-photon vibrational resonance accessible by LWIR pulses with wavelength $\lambda_L = 4\pi c / \Omega_v$: $8.4$ $\mu$m or $12.6$ $\mu$m for $N_2$ or $O_2$ respectively.

To determine the solution to Eq. (3), we express the laser pulse electric field as a plane wave modulating an envelope, $E = \hat{E}(t)\sin(\omega_L t + \phi)$. Resonant excitation requires non-impulsive excitation, such that the pulse duration, $\sigma$, far exceeds the vibrational period, $\sigma\Omega_v \gg 1$. The solution to Eq. (3) can then be expressed as $\chi_v = \chi_{v0} + \chi_{v+} + \chi_{v-}$, where

$$\chi_{v0} \simeq \frac{\gamma \hat{E}^2(t)}{2\Omega_v^2}, \quad (4)$$

$$\chi_{v+} \simeq -\frac{\gamma \hat{E}^2(t)}{4\Omega_v(\Omega_v + 2\omega_L)}\cos(2\omega_L t + 2\phi), \quad (5)$$

$$\chi_{v-} = -\frac{\gamma}{4\Omega_v}\int_{-\infty}^{t}\sin[\Omega_v t - (\Omega_v - 2\omega_L)t' + 2\phi]\hat{E}^2(t')dt', \quad (6)$$

and $\gamma = 4\pi^2\varepsilon_0\eta\mu^{-1}(\partial_Q\alpha)^2$. Equations (4-6) contribute to polarizations that oscillate either in phase or in quadrature with the laser field. The in phase, or 'refractive,' components modify the phase of the pulse during propagation, while the quadrature, or 'absorptive,' components modify the amplitude. Equation (4), for instance, results in the refractive polarization

$$P_{v0} = \frac{\gamma \hat{E}^2(t)}{2\Omega_v^2}\hat{E}(t)\sin(\omega_L t + \phi). \quad (7)$$

Equation (5) also results in a refractive polarization at $\omega_L$,

$$P_{v+} = \frac{\gamma \hat{E}^2(t)}{8\Omega_v(\Omega_v + 2\omega_L)}\hat{E}(t)[\sin(\omega_L t + \phi) - \sin(3\omega_L t + 3\phi)], \quad (8)$$

but includes an additional component oscillating at $3\omega_L$ that contributes to the usual source of 3$^{rd}$ harmonic generation in air.

Equation (6) captures the resonant response of the vibrational excitation. The resulting polarization exhibits two behaviors depending on the size of the detuning,

$\Delta = \omega_L - \frac{1}{2}\Omega_v$, relative to the pulse bandwidth, $\sigma^{-1}$. For large detuning, $\sigma|\Delta| \gg 1$, the polarization is primarily refractive. The refractive component at $\omega_L$ is

$$P_{v-} \simeq \frac{\gamma \hat{E}^2(t)}{8\Omega_v(\Omega_v - 2\omega_L)} \hat{E}(t)\left[\sin(\omega_L t + \phi) - \sin(3\omega_L t + 3\phi)\right]. \quad (9)$$

As with Eq. (8), the $3\omega_L$ component contributes to usual $3^{rd}$ harmonic generation. The coefficient, $\gamma \hat{E}^2(t)/16\Delta\Omega_v$, changes signs across the resonance, which we discuss further below. For small detuning, $\sigma|\Delta| \ll 1$, the laser frequency is near-resonant, and $\chi_{v-}$ undergoes the temporal growth characteristic of a resonantly driven harmonic oscillator. In this limit, the resulting polarization is primarily absorptive,

$$P_{v-} \simeq -\frac{\gamma}{8\Omega_v}\left(\int_{-\infty}^{t} \hat{E}^2(t')\,dt'\right)\hat{E}(t)\left[\cos(\omega_L t + \phi) - \cos(3\omega_L t + 3\phi)\right], \quad (10)$$

and scales with the accumulated laser fluence, in contrast to Eqs. (7-9) which scale with $\hat{E}^2(t)$. The $3\omega_L$ component of Eq. (10) provides a novel source for $3^{rd}$ harmonic generation far more efficient than that provided by Eqs. (8) and (9). As we will see, this term plays an important role in the self-consistent evolution of the laser pulse and vibrational response.

The effective complex susceptibilities at $\omega_L$ can be extracted from Eqs. (7-10). In particular, we write $\langle\chi_j\rangle = \langle\chi_j^R\rangle + i\langle\chi_j^A\rangle$, where $j = e, r,$ or $v$, $\langle\chi_j^R\rangle \equiv [\![\sin(\omega_L t + \phi)P_j]\!]/[\![\sin(\omega_L t + \phi)E]\!]$, $\langle\chi_j^A\rangle \equiv -[\![\cos(\omega_L t + \phi)P_j]\!]/[\![\sin(\omega_L t + \phi)E]\!]$, $[\![\ ]\!]$ denotes a cycle average, and the superscripts $R$ and $A$ refer to refractive and absorptive respectively. In the refractive limit $\sigma|\Delta| \gg 1$,

$$\langle\chi_v\rangle = \frac{\gamma \hat{E}^2(t)}{2\Omega_v^2} + \frac{\gamma \hat{E}^2(t)}{8\Omega_v(\Omega_v + 2\omega_L)} + \frac{\gamma \hat{E}^2(t)}{8\Omega_v(\Omega_v - 2\omega_L)}. \quad (11)$$

Equation (11) is the standard vibrational response in the near-infrared frequency range, $\omega_L \gg \Omega_v$. The first term, $\gamma \hat{E}^2(t)/2\Omega_v^2$, tracks the pulse intensity and contributes to the instantaneous Kerr nonlinearity in air. To avoid double counting this contribution in the total polarization, the $n_2$ values measured in Refs. [7,17], which include both the electronic and vibrational contributions, have been adjusted in Table 1 to only include the electronic contribution, $n_2 \to n_2 - (3\pi^2\eta/c\Omega_v^2\mu)(\partial\alpha/\partial Q)^2$. The second and third terms in Eq. (11) contribute minimally for near-infrared frequencies.

At LWIR frequencies, the third term, the resonant vibrational response, becomes the dominant refractive contribution. Just above resonance, this term is negative, $\sim -\gamma \hat{E}^2(t)/16\Omega_v\Delta$. One might suppose that with a small enough detuning the second term could negate the nonlinear rotational and electronic responses. In practice, however, the minimum is limited to $\sim -\gamma \hat{E}^2(t)\sigma/16\Omega_v$ before the response becomes largely absorptive. In the absorptive limit, $\sigma|\Delta|\ll 1$,

$$\langle \chi_v \rangle = \frac{\gamma \hat{E}^2(t)}{2\Omega_v^2} + \frac{\gamma \hat{E}^2(t)}{8\Omega_v(\Omega_v + 2\omega_L)} + i\frac{\gamma}{8\Omega_v}\left(\int_{-\infty}^{t} \hat{E}^2(t')dt'\right), \quad (12)$$

with the last term providing the largest, by magnitude, contribution; recall, $\sigma\Omega_v \gg 1$.

## IV. Near-Resonant LWIR Propagation

To simulate the resonant two-photon vibrational excitation and its effect on propagation, we use the 1D scalar unidirectional pulse propagation equation (UPPE) [23,24]. The UPPE equation evolves each frequency component of the pulse independently. This avoids slowly varying envelope approximations, making it ideal for situations in which harmonic generation and high-order dispersion are important. Backwards propagation is, however, neglected. In a frame co-propagating with the laser pulse along the Cartesian z-axis, the transverse electric field evolves according to

$$\frac{\partial}{\partial z}\hat{E}(z,\omega) = ik(\omega)\left[1 - \frac{c}{n(\omega)v_f}\right]\hat{E}(z,\omega) + \frac{i\omega}{2c\varepsilon_0 n(\omega)}\hat{P}(z,\omega) \quad (13)$$

where "^" indicates a frequency domain quantity, $\omega$ is the conjugate variable to the moving frame coordinate $\tau = t - z/v_f$, $v_f$ is the frame velocity, $k(\omega) = n(\omega)\omega/c$, $n(\omega)$ is the linear refractive index, and $P$ is the nonlinear polarization calculated with Eqs. (1-3). We choose the frame velocity equal to the group velocity at the carrier frequency $\omega_L$: $v_f = c/[n + \omega\partial_\omega n]|_{\omega_L}$, where $n(\omega)$ is calculated using an empirical formula for air provided in Ref. [25]. Appendix 1 details the validity conditions for the 1D propagation model.

In the simulations, the temporal pulse profile was initialized as $E(0,\tau) = E_L \sin(\omega_L \tau)e^{-\tau^2/\sigma^2}$, where $\sigma = 850$ fs corresponds to a 1 ps intensity FWHM duration. The amplitude, $E_L$, was chosen to give a peak intensity, $I = \frac{1}{2}c\varepsilon_0 E_L^2$, of $1\times 10^{12}$ W/cm$^2$. For LWIR wavelengths, this intensity results in minimal ionization, justifying the absence of a free electron current in Eq. (13). Specifically, the fractional ionization is expected to be less than $10^{-10}$ based on the ionization rate presented in Ref. [26] with parameters from Ref. [27].

Figure 1 displays the effective susceptibilities as a function of time after 15 cm of propagation. Examples of below-resonance, $\lambda_L = 8.52$ μm, resonant, $\lambda_L = 8.37$ μm, and above-resonance, $\lambda_L = 8.22$ μm, N$_2$ vibrational excitations are shown from left to right respectively. For reference, the pulse intensity profile follows the effective electronic susceptibility, $\langle \chi_e \rangle$. At each wavelength, the refractive vibrational susceptibility is smaller in magnitude than the electronic and rotational susceptibilities, and, as expected from Eq. (11), it switches signs across the resonance. On resonance, the absorptive vibrational susceptibility undergoes the rapid temporal growth characteristic of a resonantly driven harmonic oscillator, surpassing the electronic and rotational susceptibilities in amplitude. At the quantum level, the laser pulse has resonantly driven coherence between the ground and first excited vibrational state. Nevertheless, the net population in the first excited vibrational state remains low, consistent with our approximation to exclude additional vibrational states in Eq. (3). In particular, one can use the density matrix expansion [12] to find the condition for small population transfer: $(\mu \hbar \Omega_v)^{-1} [\pi \varepsilon_0 (\partial_Q \alpha) \tau E_L^2]^2 \ll 1$, which for the parameters considered here evaluates to $0.02$.

Figure 2a shows the resonantly driven, absorptive vibrational susceptibility over a 3 m propagation path. As in Fig. 1, the response grows in time, but quickly decays as the pulse propagates through space. This can be seen clearly in Fig. 2b, which displays $\langle \chi_v^A \rangle$ at $\tau = 4$ ps on the bottom horizontal axis as a function of propagation distance (shared vertical axis with Fig. 2a). Examination of Eq. (3) might lead one to believe that the spatial decay of $\langle \chi_v^A \rangle$ results from a decrease in the pulse fluence due to depletion. Surprisingly, however, the fluence remains nearly constant during the spatial decay, as demonstrated by the dashed line, top horizontal axis, in Fig. 2b.

## V. Analysis of Vibrational Response Evolution

The source of the spatial decay and the self-consistent evolution of the resonant vibrational excitation can be illustrated using a reduced, multiscale analytical model. We limit the analysis to the evolution of the laser pulse, governed by the wave equation, and the vibrational response:

$$\left(\frac{\partial^2}{\partial z^2} - \frac{2}{c}\frac{\partial^2}{\partial \tau \partial z}\right)E = \frac{1}{c^2}\frac{\partial^2}{\partial \tau^2}\chi_v E \quad (14)$$

$$\left(\frac{\partial^2}{\partial \tau^2} + \Omega_v^2\right)\chi_v = \gamma E^2. \quad (15)$$

The multiscale analysis involves expanding Eqs. (14) and (15) in time and spatial scales by writing $\partial_\tau = \partial_{\tau_0} + \varepsilon \partial_{\tau_1} + ...$, $\partial_z = \partial_{z_0} + \varepsilon \partial_{z_1} + ...$, $\chi_v(z,\tau) = \sum_n \varepsilon^n \chi_{vn}(z_0, z_1, ...; \tau_0, \tau_1, ...)$, $E(z,\tau) = \sum_n \varepsilon^n E_n(z_0, z_1, ...; \tau_0, \tau_1, ...)$, and $\gamma = \varepsilon^2 \hat{\gamma}$. As we will demonstrate, the spatial decay results from resonant 3$^{rd}$ harmonic generation, motivating our expression for the electric field

$$E_0 = A\sin(\omega_L \tau_0 + \phi) + B\sin(3\omega_L \tau_0 + 3\phi) \quad (16)$$

where $\omega_L = \tfrac{1}{2}\Omega_v$ and the dependence of $A$, $B$, and $\phi$ on $z_1$ and $\tau_1$ is implied. Consistent with Eq. (16), we set the lowest order vibrational response to zero, $\chi_{v0} = 0$.

Upon performing the expansion and keeping only resonant terms in $\chi_v$, we find the following:

$$\chi_v = -2\langle \chi_v^A \rangle \sin(\Omega_v \tau + 2\phi) \quad (17)$$

$$\frac{\partial A}{\partial z} = -\frac{\Omega_v}{4c}\langle \chi_v^A \rangle (A + B) \quad (18)$$

$$\frac{\partial B}{\partial z} = \frac{3\Omega_v}{4c}\langle \chi_v^A \rangle A, \quad (19)$$

where $\langle \chi_v^A \rangle = \gamma F(\tau) / 4\Omega_v c \varepsilon_0$, $F(\tau) = \tfrac{1}{2} c \varepsilon_0 \int_{-\infty}^{\tau} A(A - 2B) d\tau'$, and we have dropped the subscripts on $z$ and $\tau$. The source of 3$^{rd}$ harmonic generation, the RHS of Eq. (19), results from the beating of $\chi_v$ with the fundamental oscillations of the laser pulse, $\omega + \Omega_v = 3\omega$. The second RHS term of Eq. (18), $\propto \langle \chi_v^A \rangle B$, accounts for depletion of the fundamental during this process, while the first term, $\propto \langle \chi_v^A \rangle A$, accounts for depletion from vibrational excitation. A simple scaling for the spatial decay length of $\langle \chi_v^A \rangle$ can be found by deriving perturbation solutions to Eqs. (18) and (19): $A \simeq A_0 + \delta A$, $B \simeq B_0 + \delta B$, and $F = F_0 + \delta F$. The pulse starts with no initial 3$^{rd}$ harmonic content such that $B_0 = 0$. Setting $A_0 = E_L e^{-\tau^2/\sigma^2}$, we have $\delta A = -(\gamma z / 16 c^2 \varepsilon_0) A_0 F_0$, $\delta B = (3\gamma z / 16 c^2 \varepsilon_0) A_0 F_0$, $F_0(\tau) = (F_L / 2)[1 + \text{erf}(2^{1/2} \tau / \sigma)]$, and

$F_1(\tau) = -(\gamma z / 4c^2\varepsilon_0)F_0^2$, where $F_L = \frac{1}{2}c\varepsilon_0(\pi/2)^{1/2}\sigma E_L^2$ is the initial pulse fluence. Well after the pulse, the amplitude of the vibrational susceptibility is then

$$\langle \chi_v^A \rangle = \left(\frac{\gamma F_L}{4\Omega_v c\varepsilon_0}\right)\left(1 - \frac{z}{Z_d}\right), \quad (20)$$

which spatially decays over the length scale $Z_d = 4c^2\varepsilon_0 / \gamma F_L$.

Equations (17-19) and the solutions above capture several features observed in the simulations. Foremost, the laser pulse resonantly excites coherence between the ground and first excited vibrational states through a two-photon transition. This results in an absorptive vibrational susceptibility, Eq. (17), that oscillates at twice the fundamental laser frequency, with an amplitude, $\langle \chi_v^A \rangle$, that grows in time. The growth of $\langle \chi_v^A \rangle$, in turn, accelerates the 3$^{rd}$ harmonic generation, evident in the presence of $\langle \chi_v^A \rangle$ on the RHS of Eq. (19). The presence of 3$^{rd}$ harmonic opens an additional two-photon channel for resonant vibrational excitation, $3\omega_L - \omega_L = \Omega_v$. This emergent excitation channel drives the vibrational resonance exactly out of phase with the $\omega_L + \omega_L = \Omega_v$ excitation. Symbolically, the $\omega_L + \omega_L = \Omega_v$ and $3\omega_L - \omega_L = \Omega_v$ excitation channels correspond to the first, $\propto A^2$, and second, $\propto -AB$, terms in $\langle \chi_v^A \rangle \propto F \propto \int^\tau A(A-2B)d\tau'$ respectively. As the fundamental amplitude depletes, $\delta A \propto -z$, and the 3$^{rd}$ harmonic amplitude grows, $\delta B \propto z$, the vibrational susceptibility spatially decays, Eq. (20).

In support of this explanation, Fig. (3) displays the normalized fluence of the 3$^{rd}$ harmonic resulting from off and on-resonant pulses, $\lambda_L = 8.57$ $\mu$m and $8.37$ $\mu$m respectively. The fluences are normalized by the total pulse fluence, such that the value represents the fraction contributed by the 3$^{rd}$ harmonic. Consistent with the analysis above, the resonant vibrational excitation accelerates 3$^{rd}$ harmonic generation, reaching a value $>3$ times that of the off-resonant pulse after $3$ m, with a conversion efficiency of $\sim 30\%$. We note that for the parameters considered here, higher order harmonics, while present, did not reach amplitudes sufficient to significantly affect propagation or the vibrational excitation.

Even in light of Fig. (3), the interpretation offered by the multiscale analysis remains qualitative. For further validation, we simulated the propagation with Eq. (13), but included only the N$_2$ vibrational polarization density: dispersion, O$_2$ nonlinearities, and N$_2$ electronic and rotational nonlinearities were omitted. The N$_2$ density fraction was increased to 1.0 accordingly. Figure (4) compares the resulting $\langle \chi_v^A \rangle$ at $\tau = 4$ ps with that calculated from the numerical solutions to Eqs. (18) and (19) and the perturbation result, Eq. (20). The pulse parameters were identical to those above. The figure clearly exhibits agreement between the simulation and analysis, while, as expected, the perturbation result agrees only for short distances.

Figure (4) also shows $\langle \chi_v^A \rangle$ when 3$^{rd}$ harmonic generation is suppressed in the simulations (achieved by only evolving frequencies satisfying $\omega \leq 2\omega_0$). The spatial decay is less severe in this case and results solely from a decrease in the pulse fluence due to two-photon vibrational absorption. Figure (5) displays this decrease. Depletion of the pulse fluence in the presence of 3$^{rd}$ harmonic generation is also displayed. The solid and dash-dotted curves correspond to the same curves in Fig (4). With resonant 3$^{rd}$ harmonic generation, the two-photon vibrational absorption is nearly eliminated and the fluence plateaus. Without 3$^{rd}$ harmonic generation, the fluence continues to drop due to vibrational absorption. Figures (4) and (5) clearly demonstrates that the spatial decay of $\langle \chi_v^A \rangle$, results from the out of phase contribution of the $3\omega_L - \omega_L = \Omega_v$ excitation, enabled by the accelerated 3$^{rd}$ harmonic generation.

It is worth noting that this cancellation phenomenon occurs, in part, because of the weak atmospheric dispersion at LWIR and MIR wavelengths. As a comparison, the distance for phase walk-off between the fundamental and 3$^{rd}$ harmonic, $L = [n(\omega_L) - n(3\omega_L)]^{-1}(\lambda_L/6)$, of a $\lambda_L = 8.4$ $\mu$m pulse is $\sim 8$ m, while that of a $\lambda_L = 800$ nm pulse is only $L \sim 6$ mm [25]. The weak LWIR dispersion allows for the extended nonlinear interaction of the fundamental and 3$^{rd}$ harmonic.

**VI. Summary and Conclusions**

We have examined the two-photon vibrational excitation of air molecules by ultrashort LWIR laser pulses. A specific example of resonant excitation of N$_2$ with a $\lambda_L = 8.37$ $\mu$m pulse was presented. Simulations and analytical calculations demonstrated that the absorptive vibrational susceptibility undergoes temporal growth, characteristic of a resonantly driven harmonic oscillator. While the vibrational response typically contributes only a small fraction of the optical nonlinearity, the absorptive contribution surpassed both the electronic and rotational nonlinearities when driven by a 1 ps, $1\times 10^{12}$ W/cm$^2$ pulse. The temporal growth of the susceptibility was shown to accelerate 3$^{rd}$ harmonic generation, providing an additional two-photon excitation channel, $3\omega_L - \omega_L = \Omega_v$. This additional channel drives the vibrational susceptibility exactly out of phase with the original $\omega_L + \omega_L = \Omega_v$ channel, resulting in spatial decay of the absorptive vibrational response. The same effects, while not presented, occur during the excitation of O$_2$ by $\lambda_L = 12.6$ $\mu$m pulses.

**Appendix A**: **Conditions on 1D propagation**

Validity of the 1D simulation and its correspondence with potential experiments requires an initial spot size large enough that the pulse remains collimated during propagation. As a rough validity condition, we write $(Lw_0^{-1})|\partial_z w| \ll 1$, where $w$ is the $e^{-2}$ radius of a Gaussian intensity profile with initial value $w_0$ and $L$ is the propagation distance. Weak dispersion at LWIR wavelengths and the relatively small bandwidth of the pulses considered here, $\omega_0 \sigma \gg 1$, minimize spatio-temporal contributions to the spot size evolution. The spot size, therefore, evolves primarily through diffraction and self-focusing, and can be approximated by $w = w_0[1+(1-\tilde{P})\tilde{L}^2]^{1/2}$, where $\tilde{L} = L/Z_R$, $Z_R = \pi w_0^2/\lambda_L$ is the Rayleigh length, $\tilde{P} = P/P_{cr}$ is ratio of the pulse power, $P = \frac{1}{2}\pi w_0^2 I$, to the self-focusing critical power, $P_{cr} = \lambda^2/2\pi n_{2,eff}$, and $n_{2,eff}$ is an effective nonlinear refractive index. The validity condition then becomes $|1-\tilde{P}|\tilde{L}^2 \ll 1$. This condition is clearly satisfied when $\tilde{P} \simeq 1$, but this scenario requires equal power at every temporal slice in the pulse, for instance a flat top temporal profile. Instead, we exploit the limit $\tilde{P} \gg 1$, providing the condition $n_{2,eff} I (L/w_0)^2 \ll 1$. Setting $n_{2,eff} = n_{2,long}$, the adiabatic value presented in Ref. [7], $I = 1\times 10^{12}$ W/cm$^2$, $\lambda_L = 8.4$ $\mu$m, $L = 3$ m, $w_0 = 1$ cm, we have $\tilde{P} = 5.4$ and $n_{2,eff} I(L/w_0)^2 = .035$, satisfying the validity condition.

**Funding** This work was supported by the Naval Research Laboratory 6.1 Base Program. J.K.W. and H.M.M. acknowledge the support of the Air Force Office of Scientific

Research (FA95501310044), the National Science Foundation (PHY1301948) and the Army Research Office (W911NF1410372).

**Acknowledgements** The authors would like to thank M. Helle, Y.-H. Chen, and A. Stamm for fruitful discussions.

| Parameter | Reference | $N_2$ | $O_2$ |
|---|---|---|---|
| Fraction | | 0.8 | 0.2 |
| $n_2$ (m$^2$/W) | 6,16 | $7.3 \times 10^{-24}$ | $9.3 \times 10^{-24}$ |
| $\Omega_v$ (s$^{-1}$) | 12 | $4.5 \times 10^{14}$ | $3.0 \times 10^{14}$ |
| $I_M$ (Js$^2$) | 4 | $1.46 \times 10^{-46}$ | $1.9 \times 10^{-46}$ |
| $\Delta\alpha$ (m$^3$) | 6 | $6.7 \times 10^{-31}$ | $10.2 \times 10^{-31}$ |
| $\mu$ (kg) | | $1.2 \times 10^{-26}$ | $1.3 \times 10^{-26}$ |
| $\partial\alpha/\partial Q$ (m$^2$) | 19 | $1.75 \times 10^{-20}$ | $1.46 \times 10^{-20}$ |

Table 1. Parameters for nonlinear polarization

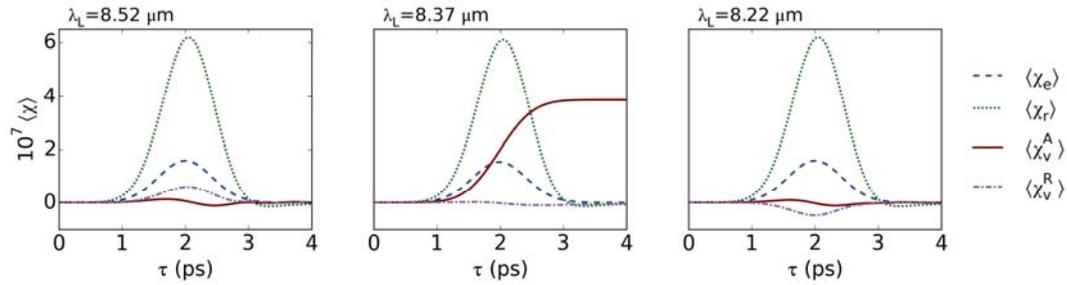

Figure 1. Amplitude of the effective electronic (blue, dashed), rotational (green, dotted), and absorptive (red, solid) and refractive (purple, dash-dot) vibrational susceptibilities as a function of pulse frame coordinate after 15 cm of propagation. From left to right the plots show examples of below-resonant, resonant, and above-resonant excitation respectively.

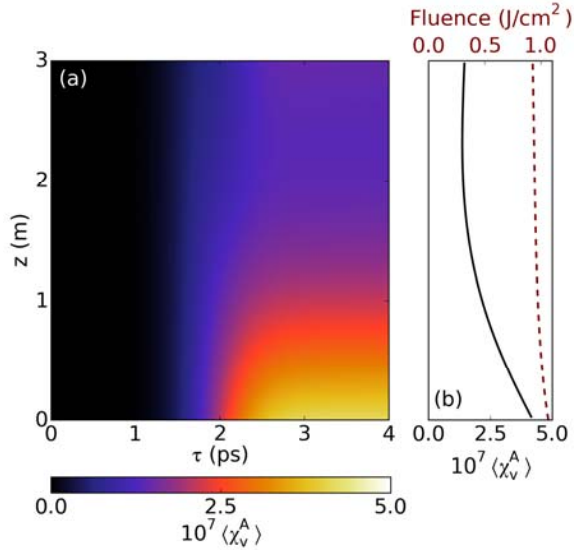

Figure 2. (a) resonant, absorptive vibrational susceptibility as a function of pulse frame coordinate and propagation distance. (b) resonant, absorptive vibrational susceptibility at $\tau = 4$ ps (black, solid, bottom horizontal axis) and total pulse fluence (red, dashed, top horizontal axis) as a function of propagation distance.

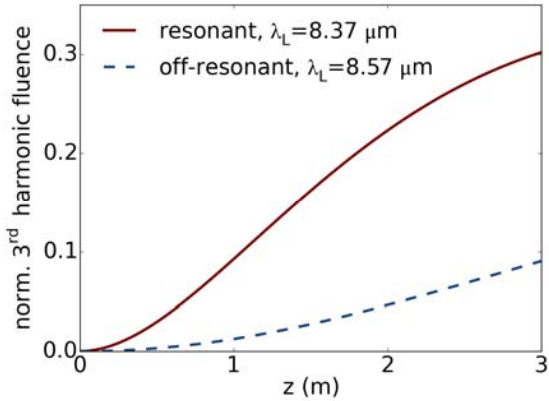

Figure 3. Normalized fluence of the 3$^{rd}$ harmonic as a function of propagation distance for resonant (red, solid) and off-resonant (blue, dashed) vibrational excitations. The values are normalized to the total fluence of the pulse.

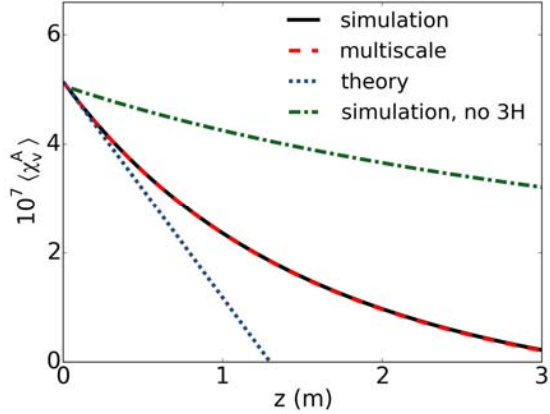

Figure 4. Resonant, absorptive vibrational susceptibility as a function of propagation distance at $\tau = 4$ ps. The solid black, red dashed, and blue dotted curves show the results from the simulation including only the $N_2$ vibrational nonlinearity, the multi-scale calculation, and the perturbation solution, Eq. (10), respectively. The green dash-dotted line displays the result of the simulations when 3rd harmonic generation is suppressed.

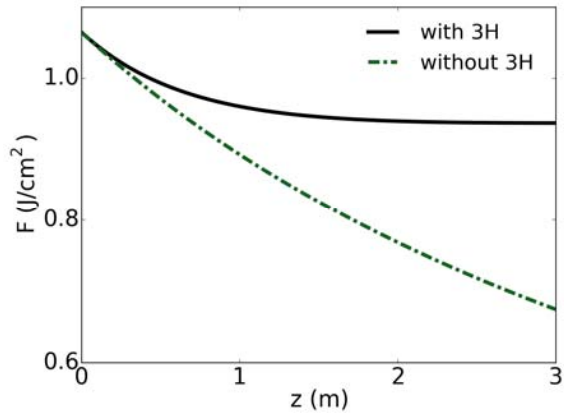

Figure 5. Pulse fluence as a function of propagation distance. The solid black and green dash-dotted lines show the results with and without 3rd harmonic generation in the simulations including only the $N_2$ vibrational nonlinearity. The resonant 3rd harmonic generation contributes to the near-elimination of two-photon vibrational absorption.

**References**
[1] P. Sprangle, J.R. Penano, and B. Hafizi, "Propagation of intense short laser pulses in the atmosphere," Phys. Rev. E **66**, 046418 (2002).